\documentclass[12pt,epsf]{iopart}
\usepackage{graphicx}
\usepackage{amssymb,latexsym}

\newcommand{\beq}{\begin{equation}}
\newcommand{\beqa}{\begin{eqnarray}}
\newcommand{\eeq}{\end{equation}}
\newcommand{\eeqa}{\end{eqnarray}}

\newcommand{\simg}{\gtrsim}
\newcommand{\siml}{\lesssim}

\def\APJ{{\it Astrophys. J. }}

\def\MNRAS{{\it Mon. Not. R. Astron. Soc. }}

\def\PLB{{\it Phys. Lett. B }}
\def\PRD{{\it Phys. Rev. D }}

\begin{document}

\title{
Classifying the Future of Universes with Dark Energy
}

\author{Takeshi Chiba}%
\address{
Department of Physics, College of Humanities and Sciences, Nihon University, 
Tokyo 156-8550, Japan}
\address{%
Division of Theoretical Astronomy, National Astronomical Observatory of Japan,
2-21-1, Osawa, 
Mitaka, Tokyo 181-8588, Japan }
\author{Ryuichi Takahashi}
\address{%
Division of Theoretical Astronomy, National Astronomical Observatory of Japan,
2-21-1, Osawa, 
Mitaka, Tokyo 181-8588, Japan }
\author{Naoshi Sugiyama}
\address{%
Division of Theoretical Astronomy, National Astronomical Observatory of Japan,
2-21-1, Osawa, 
Mitaka, Tokyo 181-8588, Japan }

\date{\today}

\pacs{98.80.H; 98.65.D}

%\preprint{astro-ph/0501661; NAOJ-Th-Ap 2005,No.xx}

\begin{abstract}
We classify the future of the universe  
for general cosmological 
models including matter and dark energy. If the equation of 
state of dark energy is less then $-1$, the age of the universe becomes finite. 
We compute the rest of the age of the universe for such universe models.  
The behaviour of the future growth of matter density perturbation is 
also studied. We find that the collapse of spherical overdensity region is greatly 
changed if the equation of state of dark energy is less than $-1$. 
\end{abstract}

\maketitle

\section{Introduction}

The Universe is replete with dark energy whose nature is almost completely 
unknown, excepting that its ``equation of state'', $w=p/\rho$, is negative. 
Dark energy determines the future of the Universe. 
Recently, there is renewed interest in the future of the Universe 
 partly because the state of the future of the universe 
would dramatically be changed in the presence of dark energy \cite{dyson,pm}. 
Moreover, another motivation comes from the fact that the 
equation of state of dark energy is being constrained by cosmological 
observations \cite{wt} and the discovery of a new type of the future 
singularity with $w<-1$ dark energy (called phantom \cite{caldwell,coy}): 
the big rip \cite{ckw}. This singularity is intriguing because a 
spacetime singularity appears even if the weak energy condition 
($\rho+p\geq 0$) is violated. 
The big rip is the singularity where the universe expands so rapidly 
due to the repulsive rather than the attractive nature of gravity 
that both the scale factor and the Hubble parameter diverge there.

In this paper, including $w<-1$ dark energy, we classify the future of 
the universe and the conformal diagrams for cosmological models with 
dark matter and dark energy with 
general equation of state. We also calculate the remaining age of the 
Universe if the age is finite. 
We also investigate the future evolution of matter density perturbation 
in the universe with dark energy.

Throughout the paper, we limit ourselves to a constant $w$. 
However, the case with time varying $w$ could be obtained  
by combining our results (however, see \cite{mc} for exceptional cases).

\section{Diagram of the Universe}

\begin{figure}
  \includegraphics[width=13cm]{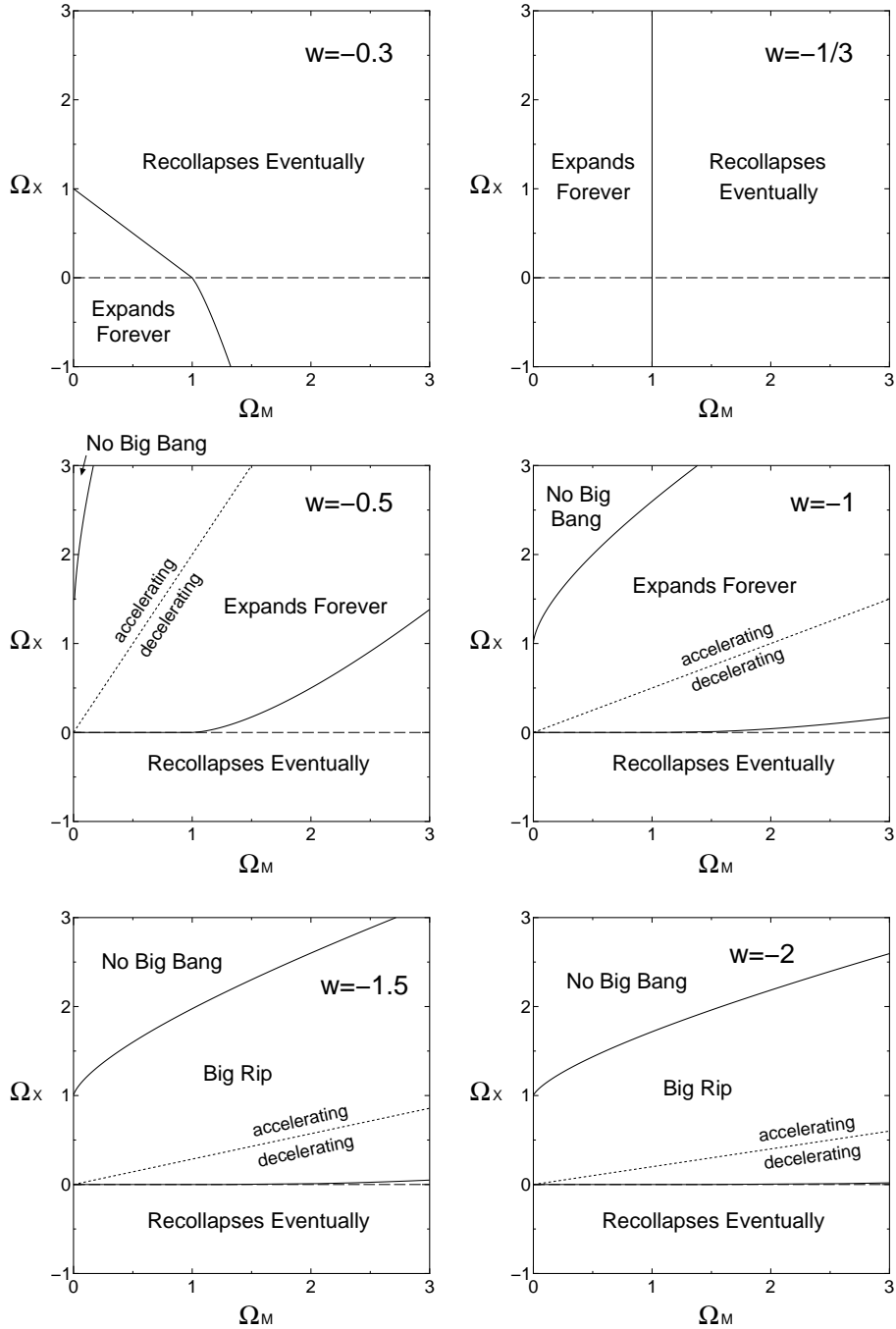}
  \caption{The ``phase diagram'' of the universe in $\Omega_M - \Omega_X$
 plane for $w=-0.3,-1/3,-0.5,-1,-1.5,-2$.
 The regions, labeled ``no big bang'', ``expands forever'', ``big rip'',
 and ``recollapses eventually'' are divided by the solid lines.
 The horizontal dashed line is $\Omega_X=0$.
 The region of ``no big bang'' represents the bounce cosmologies.
 In the region of ``expands forever'' for $w \geq -1$, 
 the universe will expand forever.
 In that of ``big rip'' for $w < -1$, the universe will end in a big rip.
 The expansion of the universe is  currently accelerating above the dotted line, 
 while decelerating below it. 
 In the region of ``recollapses eventually'', the universe will
 eventually stop its expansion and recollapse to a big crunch.
} 
\end{figure}

The Friedmann equation for the universe with (nonrelativistic) matter 
and dark energy is  
\beq
\left({H\over H_0}\right)^2=\left({\dot a\over H_0a}\right)^2=\Omega_Ma^{-3}
+\Omega_Xa^{-3(1+w)}+(1-\Omega_M-\Omega_X)a^{-2},
\label{hubble}
\eeq
where $H_0$ is the present Hubble parameter and $w$ is the equation of 
state of dark energy and $\Omega_M$ and $\Omega_X$ is 
present density parameter for matter and dark energy, respectively. 
The scale factor has been normalized such that $a=1$ at present. 

The Friedmann equation Eq.(\ref{hubble})   
can be read as the ``energy equation'':
\beq
{\dot a^2\over H_0^2}+V(a)=E,
\eeq 
where the potential energy term $V(a)$ is 
\beq
V(a)=-{\Omega_{M}}a^{-1}-\Omega_{X}a^{-1-3w},
\eeq
and the total energy $E(=\Omega_K=-K/H_0^2)$ is $E=1-\Omega_{M}-\Omega_{X}$. 
If there are roots of $E-V(a)=0$ with $a>0$, the universe can stop its
 expansion $(\dot{a}=0)$ and recollapse or bounce.
Let us discuss the roots of $E-V(a)=0$ in the following.

\paragraph*{A: $\Omega_X > 0 ~\&~ 0 > w > -1/3$. }

In this case, the dark energy behaves like a matter with small negative
 pressure.
If $E ~(=\Omega_K) < 0$, since
 $E-V(a)=1>0$ at $a=1$ and $E-V(a)\rightarrow E(<0)$ for $a \to \infty$, 
there is a root of $E-V(a)$ at $a > 1 $.  
Hence the universe will recollapse in the closed model $\Omega_K < 0$.
Interestingly, if $E > 0$ there is no root and the universe will expand forever. The future of 
the Universe would be quite different from that with a cosmological constant: 
The Universe with negative dark energy density can expand forever without big crunch. 

In top left panel of Fig.1, we show the ``phase diagram'' of the universe 
 for $w=-0.3$.
The two regions, labeled ``recollapses eventually'' and ``expands forever'',
 are divided by solid line.
The horizontal dashed line is $\Omega_X=0$.

\paragraph*{B: $w = -1/3$.}

In this special case, the Friedmann equation Eq.(\ref{hubble}) is reduced
 to $(H/H_0)^2=\Omega_M a^{-3} + (1-\Omega_M) a^{-2}$ and
 is independent of $\Omega_X$. Dark energy with $w=-1/3$ 
behaves like a curvature, so that the destiny of the 
universe depends only on the matter density. 
For $\Omega_M \leq 1$ the universe will expand 
forever, while for $\Omega_M > 1$ it will recollapse. 
In top right panel of Fig.1, we show the ``phase diagram'' for $w=-1/3$.

\paragraph*{C: $\Omega_X > 0 ~\&~ w < -1/3$. }

The universe can recollapse or bounce if $E < V_m=\max(V(a))$. 
Let $a_m$ be the scale factor at the maximum of $V(a)$, where $a_m$ is given by 
\beq
a_m=\left({\Omega_M\over -(1+3w)\Omega_X}\right)^{-1/3w},
\eeq
and $V_m$ is
\beq
V_m=-{3w\over 1+3w}{\Omega_M\over a_m}.
\eeq
Hence the critical condition $E-V_m=0$ is rewritten as,
\beq
\left| \frac{1-\Omega_M-\Omega_X}{3 ~w~ \Omega_X} \right|^{-3 w} =
 \left| \frac{\Omega_M}{(1+3 w) ~\Omega_X} \right|^{-1-3 w}.
\eeq
This equation is the same as Eq.(5) in Moles (1991) if we set $w=-1$.
The bounce would occur if $E< V_m$ and $a_m<1$;
the recollapse would occur if $E< V_m$ and $a_m>1$ \cite{moles}. 

In Fig. 1, we display the ``phase diagram'' of
 the universe depending on the equation 
 of state of dark energy ($w=-0.5,-1,-1.5,-2$).
For $w < -1/3$, the three regions are divided by the two solid lines.
The region of ``no big bang'' represents the bounce cosmologies.
As $w$ becomes negatively larger, the region of bounce 

becomes larger. This can be understood from  
\beq
{\partial V_m\over \partial w}={V_m\over w}\left(\ln a_m-
\ln(\Omega_X/\Omega_M)\right)
\eeq
for fixed $\Omega_M$ and $\Omega_X$. 
This implies that 
for $a_m<1$ (or $\Omega_M<-(1+3w)\Omega_X$), $V_m$ increases 
as $w$ decreases when $\Omega_M$ and $\Omega_X$ fixed, so that 
the universe can bounce more easily. 
On the other hand, for $a_m>1$, $V_m$ decreases as $w$ 
decreases when $\Omega_M$ and $\Omega_X$ fixed, so that the universe 
can recollapse more hardly. 

If the equation of the state of dark energy is less than $-1$ 
\cite{caldwell,coy}, then 
the universe expands  so rapidly that the scale factor will diverge 
and the space will eventually be torn apart 
and the universe will consequently result in the ``big rip'' \cite{ckw}: 
the singularity with the positively divergent Hubble parameter. 
In the bottom panels of Fig.1 $(w=-1.5,-2)$, the region of ``big rip''
 represents the cosmological models with big rip singularity.
\footnote{The properties of the singularity and the existence 
of a stable fixed point are studied in \cite{ont}.}

\paragraph*{D: $\Omega_X < 0 ~\&~ w \neq -1/3$. }

For $0 > w > -1/3$ the universe will recollapse if $E < V_m$. 
While for $w<-1/3$ there is a root at $a>1$, since $E-V(a)$ is $1$ at
 $a=1$ and $\Omega_X a^{-1-3w} ~(<0)$ at $a \gg 1$.
Hence the universe will recollapse in $\Omega_X < 0 ~\&~ w < -1/3$.

\section{The Life of the Universe}

\begin{figure}
\includegraphics[width=\hsize]{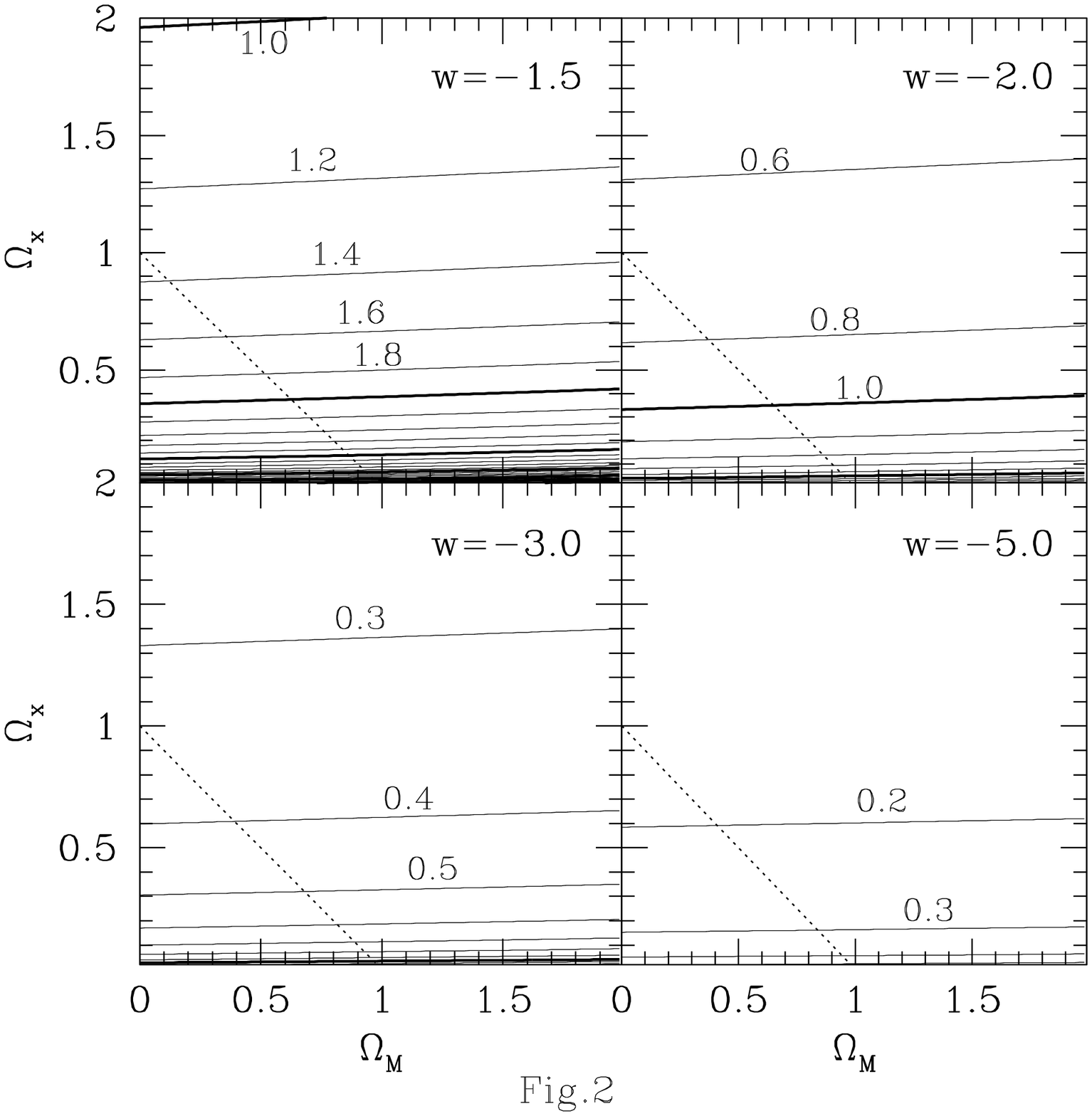}
\caption{$H_0t_{\rm left}$ for general universe model: 
for $w=-1.5$, $H_0t_{\rm left}$ is  $3.0$ from top in step of $0.2$;  
for $w=-2.0$, $0.6$ from top in step of $0.2$; for $w=-3.0$, $H_0t_{\rm left}$ is 
$0.3$ from top in step of $0.1$; for $w=-5.0$, $H_0t_{\rm left}$ is 
$0.2$ from top in step of $0.1$. 
}
\label{fig:fig2}
\end{figure}

For $w<-1$, the age of the Universe becomes finite even if the energy 
density of dark energy is positive. Then the immediate question would be, 
"how much time is left for the Universe?". 
So we calculate numerically the remaining age of 
the universe, $t_{\rm left}$,  for general non-flat universe with $w<-1$:
\beqa
t_{left}&=&\int_1^{\infty}{da\over \dot a}=\int_1^{\infty}{da\over a H}
\nonumber\\
&=&{1\over H_0}\int_0^1{dx\over 
x\sqrt{\Omega_Mx^3+\Omega_Xx^{3(1+w)}+(1-\Omega_M-\Omega_X)x^2}},
\label{tleft}
\eeqa
where we have introduced $x=1/a$. The case of flat universe is calculated in 
\cite{caldwell,cht,ss}. 

Fig. \ref{fig:fig2} is $H_0t_{\rm left}$ for general non-flat universe 
models: for $w=-1.5$, $H_0t_{\rm left}$ is  $1.0$ from top in step of $0.2$; 
for $w=-2.0$, $H_0t_{\rm left}$ is $0.6$ from top in step of $0.2$; 
for $w=-3.0$, 
$H_0t_{\rm left}$ is $0.3$ from top in step of $0.1$; 
for $w=-5.0$, $H_0t_{\rm left}$ is 
$0.2$ from top in step of $0.1$.

\begin{figure}
\includegraphics[width=\hsize]{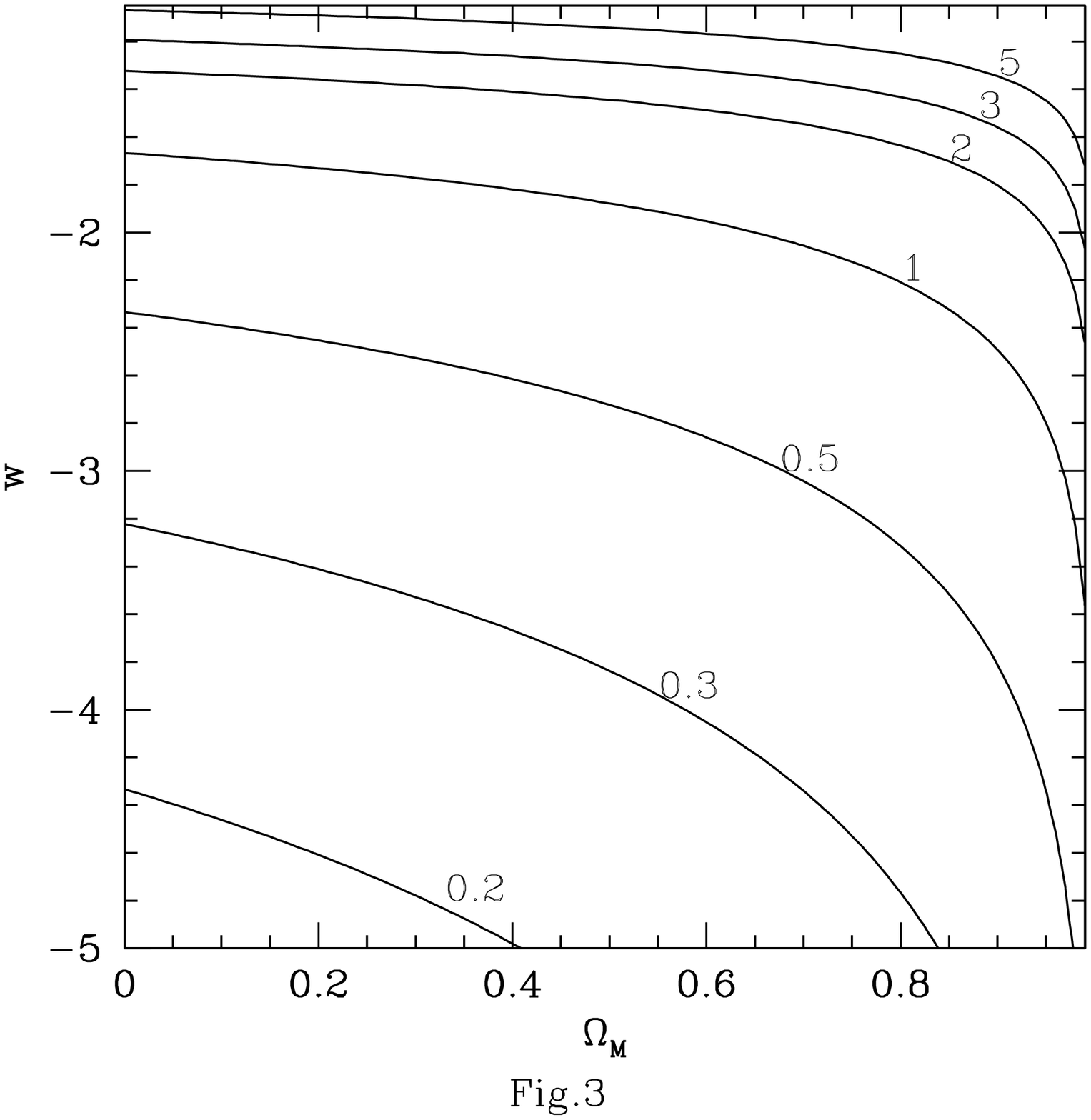}
\caption{$H_0t_{\rm left}$ for flat models. $H_0t_{left}=
5.0,3.0,2.0,1.0,0.5,0.3,0.2$ from top to bottom. 
}
\label{fig:fig3}
\end{figure}

Fig. \ref{fig:fig3} is for a flat model: $H_0t_{\rm left}$ is 
$5.0,3.0,2.0,1.0,0.5,0.3,0.2$ from top to bottom. 
{}From Fig. \ref{fig:fig2}, we find that the remaining age is 
primarily determined by $\Omega_X$. Moreover, for negatively 
larger $w$ dependence on $\Omega_X$ becomes weak and the age 
is essentially determined by $w$ (see Fig. 2 for $w=-5.0$). 
This fact can easily be found from Eq.(\ref{tleft}); 
since for $|w|\gg 1$ dark energy becomes dominant during most of 
the remaining age of the universe, the integral can be approximately 
estimated as
\beqa
H_0t_{left}\simeq \int^1_0{dx\over \sqrt{\Omega_X}x^{(5+3w)/2}}=
{2\over 3|1+w|\sqrt{\Omega_X}},
\eeqa
which explicitely shows that $t_{left}$ is more sensitive to $w$ than to 
$\Omega_X$. It is found from numerical calculations that 
this approximates the integral Eq.(\ref{tleft}) well (within 10\%) for 
$0.5\siml \Omega_X\siml 2.0$ if $w\siml -1.3$. 

\section{Conformal Diagrams}

Similar to the fate of the universe, an interesting question 
would be ``what is the entire structure of spacetime for the 
universe with dark energy?''.  
We show the conformal diagrams of flat cosmological models with
 various dark energy. There are several 
 works which have some overlap \cite{fst,alex,hst,gb,loeb,hst,sorbo}. 
Our aim here is to collect and classify these results 
including matter and $w<-1$ case for comparison and for completeness. 

\subsection{Dark Energy without Matter}
For simplicity, we first consider cosmological models without matter. 
The conformal diagrams of such cosmological models (for $w\geq -1$) are 
studied in \cite{fst,hks}. We include cosmological models with $w<-1$ 
dark energy as well. 

Since the metric  of flat models is conformal to Minkowski spacetime
\beq
ds^2=-dt^2+a(t)^2(dr^2+r^2d\Omega^2)=a(\eta)^2(-d\eta^2+dr^2+r^2d\Omega^2),
\eeq
the conformal diagram is a subset of that of the Minkowski spacetime and 
the range of the conformal time $\eta$ depends on $w$. 

\paragraph*{A: $w>-1/3$. } 

Since $a(t)=t^{2/3(1+w)}$, 
\beq
\eta=\int {dt\over a}= {3(1+w)\over 1+3w}t^{(1+3w)/3(1+w)}.
\eeq
The range of $\eta$ corresponding to that of $t$ ($0\leq t< \infty$) is, 
$0\leq \eta < \infty$, so that the conformal diagram is the upper half 
of the conformal diagram of Minkowski spacetime. 

%\paragraph*
\paragraph*{B: $w=-1/3$. } 

Since $a=t$, $\eta = \ln t$. The range of $\eta$ is 
$-\infty <\eta < \infty$, and the diagram is almost the same 
as that of the Minkowski spacetime excepting  the presence  
of big-bang singularity at the past null infinity. 

Since there is a past null singularity, the physical size of the past 
light cone $d_H$ for an observer at $r=0$ (particle horizon) is given by
\beq
d_H=a(t)\int_0^t{dt'\over a(t')},
\eeq
and it is infinite for $w=-1/3$. Hence, there is no horizon problem 
in this case.

%\paragraph*
\paragraph*{C: $-1<w<-1/3$.}

Since $a(t)=t^{2/3(1+w)}$ and the exponent is greater than unity, 
$\eta \propto -t^{(1+3w)/3(1+w)}$. The range of $\eta$ corresponding to 
that of $t$ ($0\leq t< \infty$) is $-\infty< \eta \leq 0$, 
so that the conformal diagram is the lower half of the conformal diagram 
of Minkowski spacetime with the big-bang singularity. 
Since there is a past null singularity, $d_H$ is infinite 
in this case as well.

%\paragraph*
\paragraph*{D: $w=-1$. }

Since $a=\exp(Ht)$ $(-\infty < t < \infty)$, $\eta\propto -\exp(-Ht)$. 
Again, the conformal diagram 
is the lower half of the conformal diagram of Minkowski spacetime 
but without the big bang singularity. 

%\paragraph*
\paragraph*{E: $w<-1$.}

Since $a=(-t)^{2/3(1+w)}$, $\eta \propto -(-t)^{(1+3w)/3(1+w)}$. 
The range of $\eta$ corresponding to that of 
$t$ ($-\infty < t\leq 0$) is $-\infty< \eta \leq 0$, 
so that the conformal diagram is again the lower half of the conformal 
diagram of Minkowski spacetime but this time with the big rip singularity 
at the future spacelike infinity. Since the past null singularity is null, 
$d_H$ is infinite in this case as well.

\begin{figure}
\includegraphics[width=\hsize]{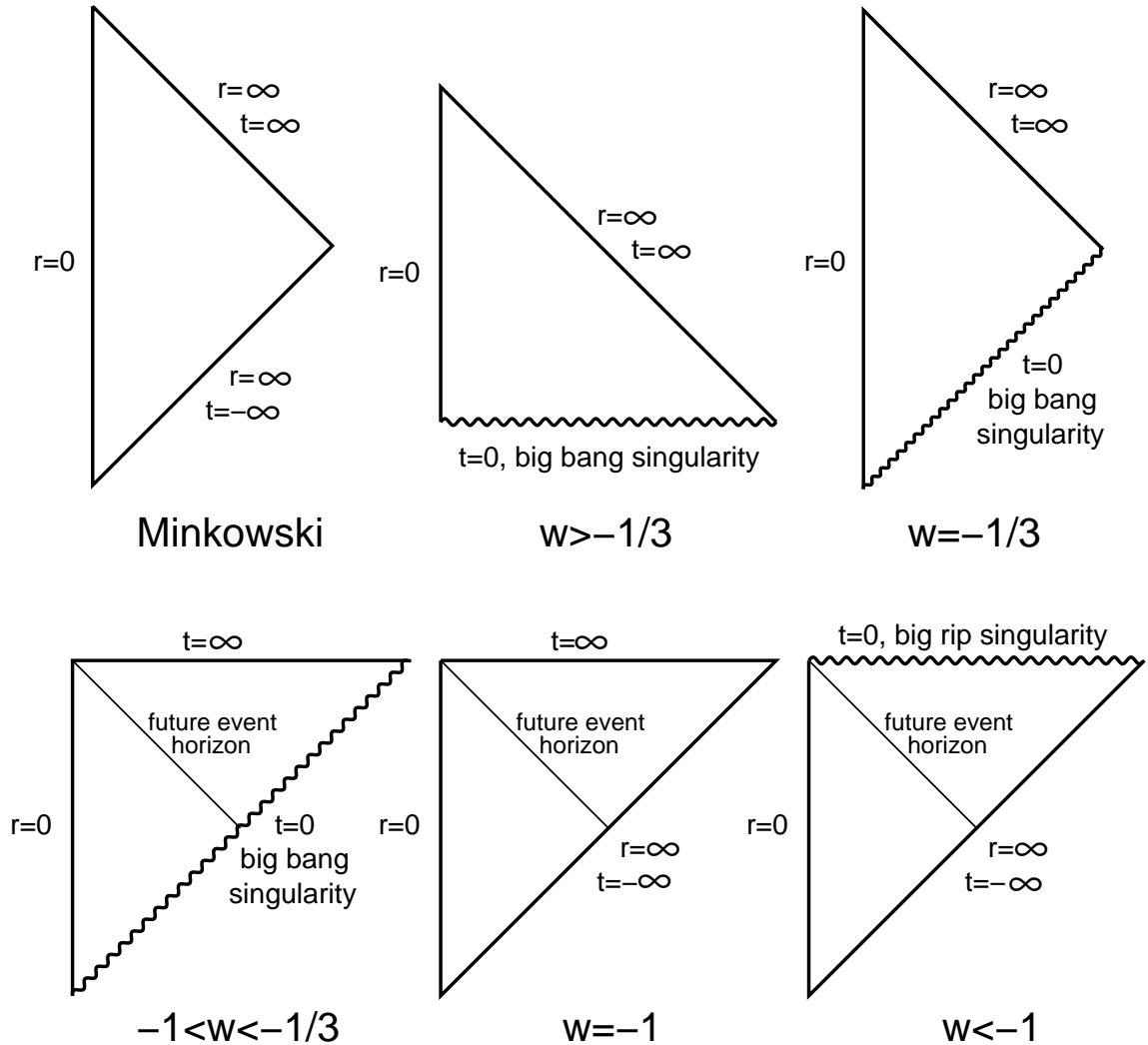}
\caption{Conformal diagrams of flat cosmological models with dark energy 
but without matter for $w>-1/3,w=-1/3,-1<w<-1/3,w=-1,w<-1$.
The thin solid line is the future event horizon.
}
\label{fig:diagram-matter}
\end{figure}

These results are shown in Fig. \ref{fig:diagram-matter}.
For $w<-1/3$, since the conformal time is bounded above, there is a 
future cosmological event horizon \cite{gw,hks}. The light rays 
emitted beyond the horizon never reach an observer at $r=0$. 
Hence, the asymptotic region of spacetime cannot be measured, and there is 
no S-matrix \cite{hks}.\footnote{However, the situation is not improved 
much for $w>-1/3$. see \cite{bousso}.} 
The proper radius of the horizon $R_c$ is given by
\beq
R_c=a(t)\int^{\infty}_t{dt'\over a(t')}=-{3(1+w)\over 1+3w}t,
\eeq
for $-1<w<-1/3$. $R_c=H^{-1}$ for $w=-1$, and $R_c=(3(1+w)/(1+3w))(-t)$ 
for $w<-1$. (Note that the range of $t$ is bounded above for $w<-1$.) 
The size of the horizon grows for $-1<w<-1/3$ and remains constant for 
$w=-1$ but {\it decreases} for $w<-1$ because phantom matter violates the 
dominant energy condition. 

\begin{figure}
\includegraphics[width=\hsize]{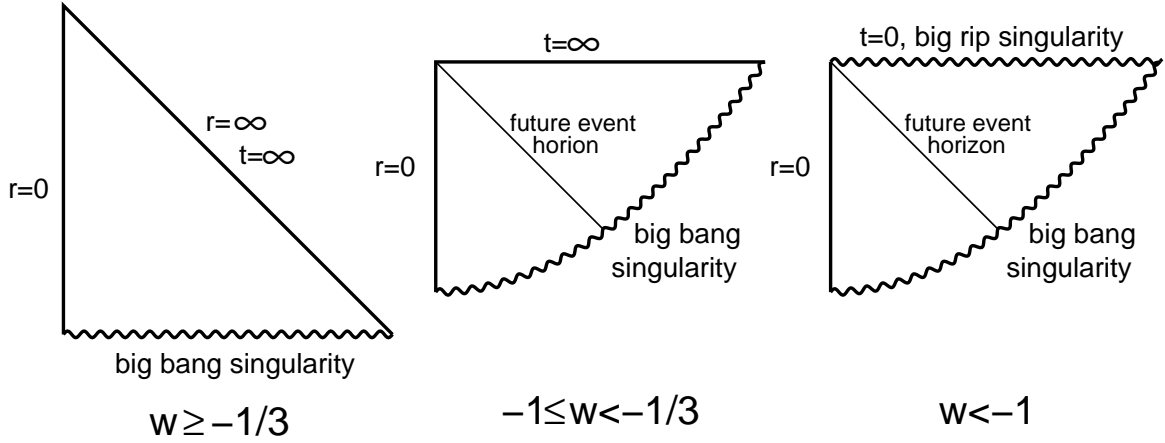}
\caption{Conformal diagrams of flat cosmological models with matter 
and dark energy for $w\geq -1/3, -1\leq w<-1/3, w<-1$. 
The thin solid line is the future event horizon.
}
\label{fig:diagram+matter}
\end{figure}

\subsection{Including Matter}

When matter is included, lower half parts of the diagrams 
are replaced with big bang singularity. 
We plot the corresponding diagrams in Fig. \ref{fig:diagram+matter}. 
Because matter was dominated in the past, ]
the size of the particle horizon 
becomes finite even for $w\leq -1/3$. 
However, the presence of the cosmological event horizon 
is not affected by including matter, since the universe 
would be dominated by dark energy in the future 
 (see \cite{gb,loeb,hst,sorbo} for the detailed 
discussion of the evolution of the horizon).

\section{Growth of Structures}

In this section, we study the formation of structures in 
flat universe models with dark energy. First, we consider 
the evolution of linear matter perturbation, 
and then on the base of it, we consider the nonlinear evolution 
of perturbation using spherical collapse model. 

\subsection{Linear perturbation}

The growth of linear matter density perturbation 
$\delta=\delta\rho_M/\rho_M$ in various flat cosmological models with 
dark energy is determined by the following equation \cite{peebles}:
\beq
\ddot{\delta}+2H\dot{\delta}-4 \pi G \rho_M \delta=0.
\label{linear}
\eeq
Eq.(\ref{linear}) can be rewritten in terms of $g = \delta/a$ as %%%%%
\beq
(\Omega_M+\Omega_xa^{-3w}){d^2g\over d\ln a^2}+
\left({5\over 2}\Omega_M+{5-3w\over 2}\Omega_Xa^{-3w}\right){dg\over d\ln a}
+{3\over 2}(1-w)\Omega_Xa^{-3w}g=0.
\label{lindel}
\eeq 

\begin{figure}
\includegraphics[width=\hsize]{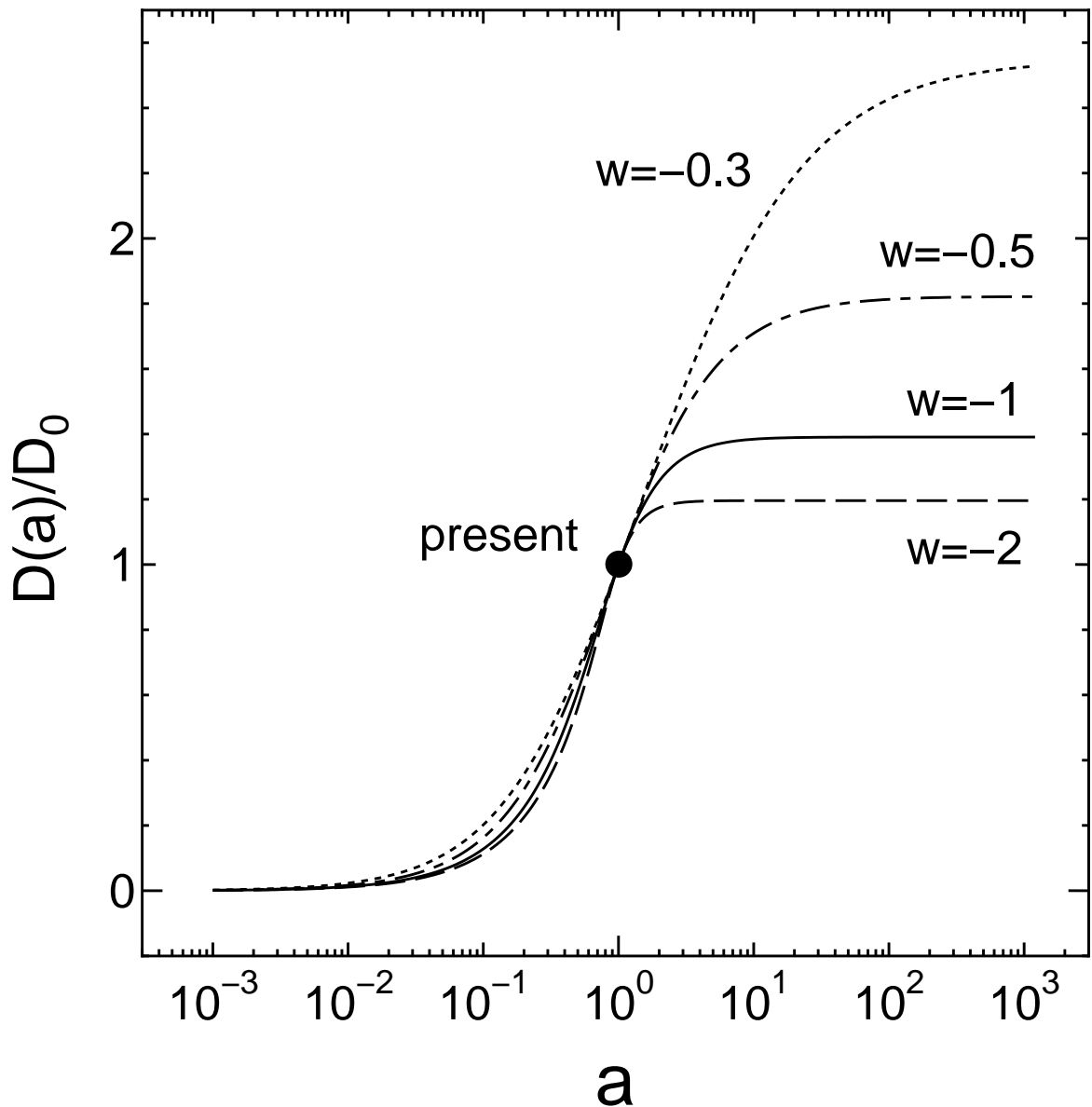}
\caption{
The linear growth rate as a function of scale factor for $w=-0.3$(dot),
 $-0.5$(dot-dash), $-1$(solid) and $-2$(dash) with $\Omega_M=0.3$. 
$D(a)$ is normalized by its present value, $D_0 \equiv D(a=1)$. }
\label{fig:growth-a}
\end{figure}

\begin{figure}
\includegraphics[width=\hsize]{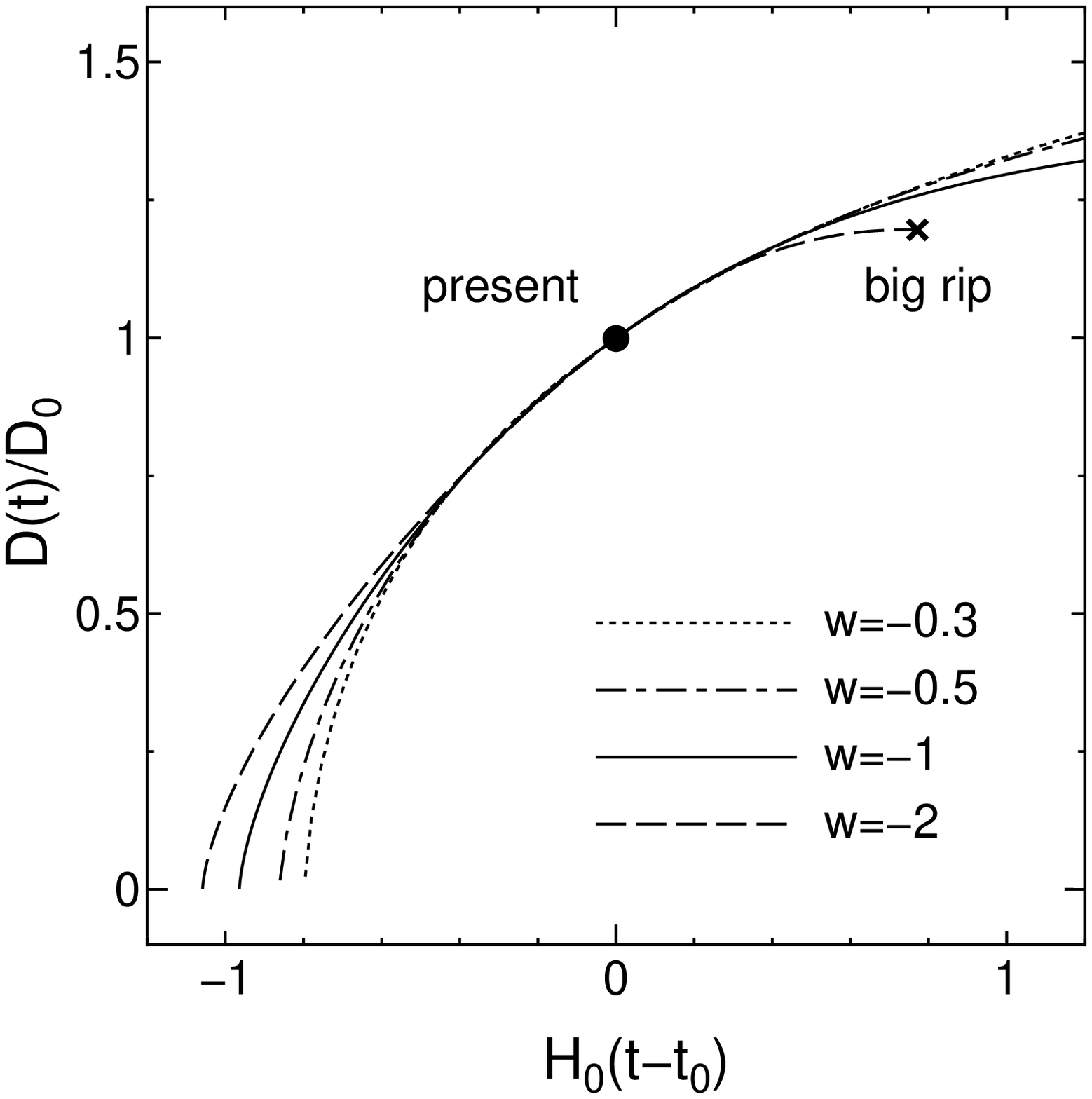}
\caption{
Same as Fig.\ref{fig:growth-a}, but as a function of cosmic time.
For $w=-2$, the dashed line ends in big rip singularity (cross).
}
\label{fig:growth-t}
\end{figure}

The exact solution of the linear perturbation equation (\ref{lindel})
 was obtained by Silveira and Waga \cite{sw,pad03} 
(see also \cite{lh} for $w=-1$) 
for a constant $w$. Denoting the solution as $D(a)/a$, it is given by 
\beq
  \frac{D(a)}{a} = {}_2F_1 \left( -\frac{1}{3w}, \frac{w-1}{2w},
 1-\frac{5}{6w}; -\frac{\Omega_X}{\Omega_M} a^{-3w} \right),
\label{lgr}
\eeq
where ${}_2F_1$ is the Gauss's hypergeometric function.  
Here ${D(a)}$ is normalized so that ${D(a)} \to a$ at
 $a \to 0$.

We compute the evolution of density perturbation as a function of scale 
factor or cosmic time. The results are shown in Fig. \ref{fig:growth-a} and 
Fig. \ref{fig:growth-t}. 
Matter density perturbation is normalized by its present value,
 $D_0 \equiv D(a=1)$. 
We assume $\Omega_M=1-\Omega_X=0.3$.
The future growth of the density perturbation becomes more suppressed 
for negatively large $w$. 
%Although the linear growth rate $D$ is higher in the past for 
%negatively larger $w$, $D$ is saturated earlier in the 
%future due to the rapid growth of dark energy density relative to matter 
%density, and consequently $D$ will be terminated toward a lower value.
We find from Eq.(\ref{lgr}) that $D(a)$ asymptotically converges toward 
a constant value in the limit of $a \to \infty$ for $w < -1/3$ \cite{as}, %%%%%
\beqa
  D(a) &\rightarrow& \frac{\Gamma(1-\frac{5}{6w}) \Gamma(\frac{1}{2}
 -\frac{1}{6w})}{\Gamma(1-\frac{1}{2w}) \Gamma(\frac{1}{2}-\frac{1}{2w})}
 \left( \frac{\Omega_X}{\Omega_M} \right)^{1/3w}\Bigl{(}1+\nonumber\\
&&~~~~~~~\frac{\Gamma(-\frac{1}{2}+\frac{1}{6w})\Gamma(1-\frac{1}{2w})
\Gamma(\frac{1}{2}-\frac{1}{2w})}
{\Gamma(\frac{1}{2}-\frac{1}{6w})\Gamma(\frac{1}{2}-\frac{1}{3w})\Gamma(-\frac{1}{3w})}
\left(\frac{\Omega_X}{\Omega_M} \right)^{\frac{1}{6w}-\frac{1}{2}}
a^{-\frac{1}{2}+\frac{3w}{2}}\Bigr{)},
\label{dinfty1}
\eeqa
while for $0>w>-1/3$
\beq
  D(a) \rightarrow \frac{\Gamma(1-\frac{5}{6w}) \Gamma(\frac{1}{2}
 -\frac{1}{6w})}{\Gamma(1-\frac{1}{2w}) \Gamma(\frac{1}{2}-\frac{1}{2w})}
 \left( \frac{\Omega_X}{\Omega_M} \right)^{1/3w}\left(1+\frac{w-1}{w(6w-5)}
\left(\frac{\Omega_M}{\Omega_X}\right)a^{3w}\right).
\label{dinfty2}
\eeq
This is because the time scale for expansion $\sim H^{-1} \sim
 (G \rho_X)^{-1/2}$ is much less than that for growth of density
 perturbation $\sim (G \rho_M)^{-1/2}$ in the future ($\rho_X \gg \rho_M$)
 \cite{peebles}.  The asymptotic constant in Eq.(\ref{dinfty1}) and 
E.(\ref{dinfty2}) is the same for both cases and is an increasing 
function of $w$. 
Eq.(\ref{dinfty1}) and Fig.\ref{fig:growth-a} show that for negatively %%%%%
larger $w$, the growth of $D$ is saturated earlier in the future. 
This is due to the rapid growth of $\rho_X$ relative to $\rho_M$.
The difference of $D$ between various $w$ may be suppressed when 
$D$ is plotted as a function of cosmic time since time elapses more 
slowly for negatively large $w$.

\subsection{Spherical collapse}

In the previous section we studied the growth of linear density
 perturbation.
In this section we consider spherical collapse model   
 in order to investigate nonlinear growth of perturbation. 
It is already studied both analytically and numerically \cite{nl} 
for a cosmological constant and the fate of structures for $w<-1$ is 
briefly described in \cite{ckw}. Our approach here will be more analytic than the 
former and more quantitative than the latter for general 
equation of state of dark energy. 
 
We consider a evolution of spherical overdense region with uniform
 density $\rho_{sp}$ and radius $R$ \cite{ws98}.
The evolution of radius is described by
\beq
 \frac{\ddot{R}}{R} = - \frac{4 \pi G}{3} \rho_{sp} -
  \frac{4 \pi G}{3} \left( 1+3w \right) \rho_{X},
\label{eom}
\eeq
where $\rho_{sp} \propto R^{-3}$ and
 the mass is $M=(4 \pi \rho_{sp}/3) R^3$. 
The first term represents the gravitational force which contracts the
 sphere, while the second term represents the repulsive force
 for $w < -1/3$  due to the dark energy and it prevents the contraction.

Defining dimensionless radius $y \equiv R/R_0$, where $R_0$ is the radius
 at present, the above equation (\ref{eom}) is rewritten as, 
\beqa
 \left( \Omega_M + \Omega_X a^{-3w} \right) \frac{d^2}{d \ln a^2}
 \left( \frac{y}{a} \right) &+& \frac{1}{2} \left[ \Omega_M +
 \left( 1-3w \right) \Omega_X a^{-3w} \right] \frac{d}{d \ln a}
 \left( \frac{y}{a} \right) \nonumber \\
 && - \frac{1}{2} \Omega_M \frac{y}{a}
 + \frac{1}{2} \Delta_0 \Omega_M \left( \frac{y}{a} \right)^{-2}=0,
\label{eom2}
\eeqa
where $\Delta_0 \equiv \rho_{sp}(R_0) / \rho_{M,0}$ is the 
 ratio of sphere to background density at present and it represents
 nonlinear density contrast.
We also define linear density contrast $\delta_0$ which is the density
 contrast at present if the density perturbation evolve by linear growth
 rate $D(a)$.
%Since the perturbation is linear at early time $a \ll 1$,
Then, $\delta_0$ is given by
\beq
 \delta_0 = \lim_{a \to 0} \frac{\delta(a)}{D(a)} D_0 = \lim_{a \to 0}
 \frac{1}{D(a)} \left[ \frac{\rho_{sp}(R(a))}{\rho_M(a)} -1 \right] D_0.
\eeq
The analytical solution of Eq.(\ref{eom2}) at $a \ll 1$ is
 obtained by
\beq
  \frac{y}{a}=\Delta_0^{1/3} \left[ 1 - \frac{1}{3} \frac{\delta_0}{D_0}
 a + \mathcal{O}(a^2) \right].
\eeq 
We use the above equation as the boundary condition to solve
 Eq.(\ref{eom2}).
Either $\Delta_0$ or $\delta_0$ can be determined by a
 condition of $y(a=1)=1$, and hence there is one free parameter.

In order to relate the mass $M$ of sphere to the density contrast
 $\delta_0$,
 we assume that the overdense region is formed from 
 the $1 \sigma$ high-density peak of mass fluctuation.
Then, the linear perturbation at present $\delta_0$ is equal to
 the linear mass fluctuation $\delta_M$ for the mass $M$,
 $\delta_0=\delta_M$.
% and hence the mass $M$ is related to $\delta_0$ or $\Delta_0$.
Here, $\delta_M$ is given by
\beq
  \delta_M^2 = \frac{1}{2 \pi^2} \int dk k^2 P(k) W^2(kr),
\label{delm}
\eeq  
where $P(k)$ is the power spectrum and
 $W(kr)$ with $r=(2M/\Omega_M H_0^2)^{1/3}$ is the top-hat window
 function \cite{bar86}.
We normalise $\delta_M$ so that $\delta_M=\sigma_8=0.9$ at
 $r=8 h^{-1}$ Mpc.
We assume $\Omega_M=0.3$, the baryon density $\Omega_b=0.04$ and
 the Hubble parameter $h=0.7$.

\begin{figure}
\includegraphics[width=\hsize]{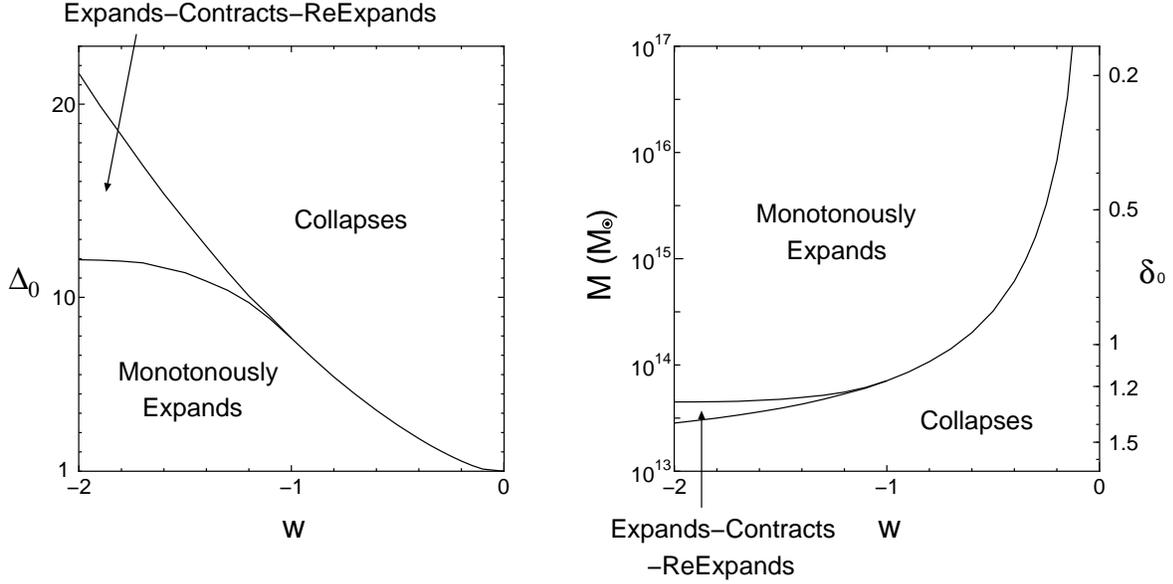}
\caption{
The fate of spherical overdense region.
$\Delta_0$ is the ratio of spherical density to background density at
 present, $M$ is its mass and $\delta_0$ is the linear perturbation 
 at present.
In the region of ``Monotonously Expands'', the sphere monotonously expands
 forever.
In the region of ``Collapses'', the sphere stops its
 expansion and collapses to $R=0$.
In the region of ``Expands-Contracts-ReExpands'', the sphere stops its
 expansion and turns around. But the repulsive force due to the dark energy,
 which increases with time,
 prevents its contraction and the sphere re-expands forever.
}
\label{sc2627}
\end{figure}

Evaluating the dimensionless radius $y(a)$ in Eq.(\ref{eom2}) for various
 values of parameters $\Delta_0$, $M$ and $\delta_0$,
 we find that the fate of over-density sphere can be classified into 
three cases. 
We display these cases in Fig.\ref{sc2627}. 
The left panel is $\Delta_0-w$ plane, while the right panel is 
 $M-w$ plane (we also show $\delta_0$).

\begin{enumerate}

\item The region of ``Monotonously Expands''.
 For low density region, $\Delta_0 \siml 10$,
 the sphere monotonously expands forever.
 This is because  the linear perturbation at far future,
 $\lim_{a \to \infty} \delta(a) = \delta_0/D_0 \lim_{a \to \infty} D(a)$, 
 is a constant value from Eq.(\ref{dinfty1}) and Eq.(\ref{dinfty2}) and
 cannot reach the critical over density $\delta_c \sim 1.68$ \cite{ps74}
 which is $\delta$ when the sphere collapses. 
 In the far future, $a \gg 1$, the radius increases in proportional to
 the scale factor, $R \propto a$. 

\item The region of ``Collapses''.
 For high density region, the sphere stops its
 expansion and collapses to $R=0$.

\item The region of ``Expands-Contracts-ReExpands''.
 For intermediate mass density, $10 \siml \Delta_0 \siml 20$ with $w < -1$,
 the sphere stop its expansion and turns around.
 But the repulsive force due to the dark energy, which increases with time,
 prevents its contraction and the sphere re-expands forever.  

\end{enumerate}

\begin{figure}
\includegraphics[width=\hsize]{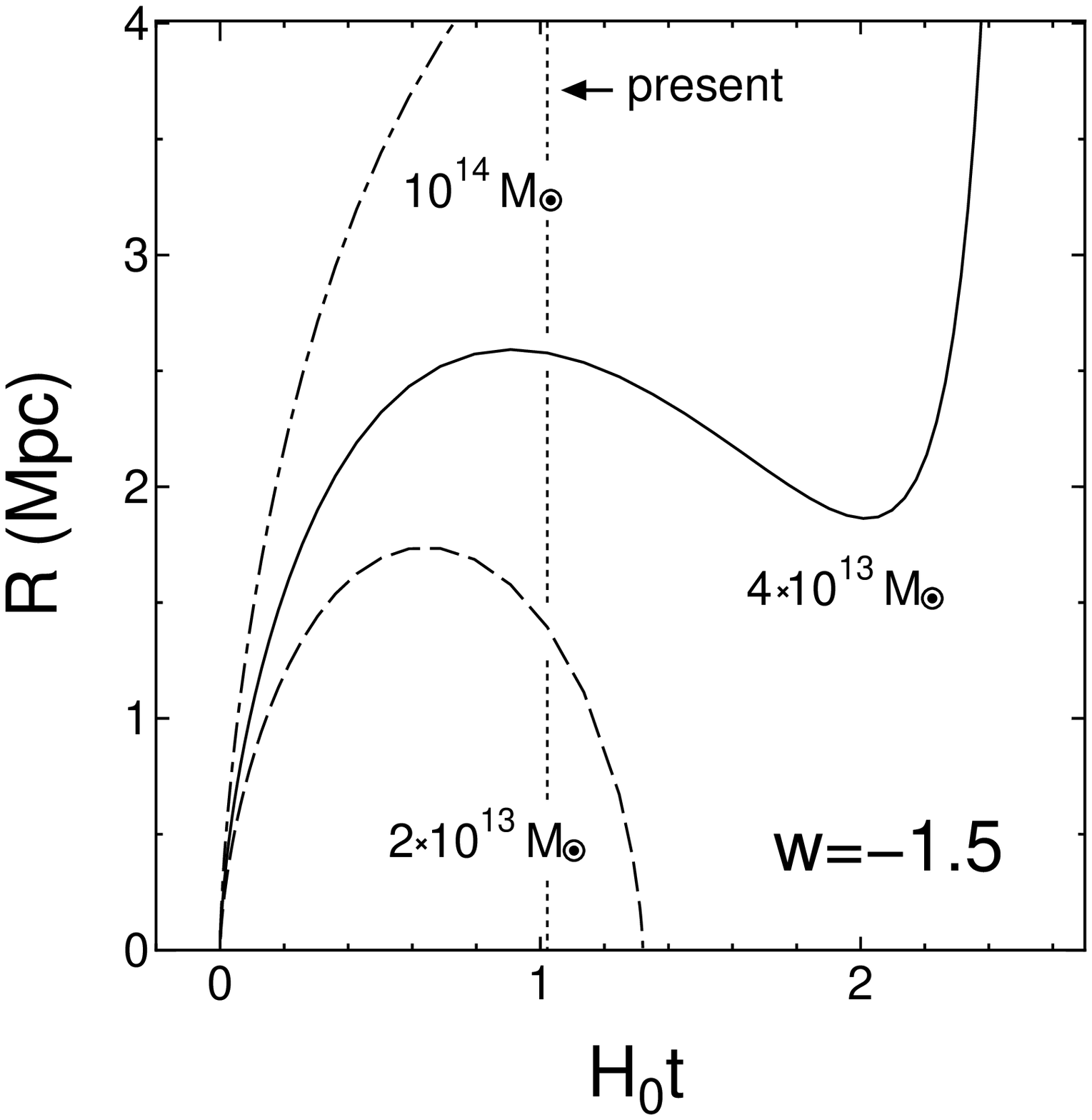}
\caption{
The radius $R$ as a function of the cosmic time with $w = -1.5$ for
 $M=10^{14} M_\odot$ (dot-dash), $4 \times 10^{13} M_\odot$ (solid)
 and $2 \times 10^{13} M_\odot$ (dash).}
\label{sc7}
\end{figure}

In Fig.\ref{sc7}, we show the radius $R$ as a function of time with $w=-1.5$
 for $M=10^{14} M_\odot$ (dot-dash), $4 \times 10^{13} M_\odot$ (solid)
 and $2 \times 10^{13} M_\odot$ (dash), as an example of these three types.
The radius at present is $R_0=[3 M/(4 \pi \rho_{M,0} \Delta_0)]^{1/3}$.
As shown in Fig.\ref{sc7}, we note that the behavior of $R$ is not
 symmetric about the turn-around time for $w \neq -1$.

\begin{figure}
\includegraphics[width=\hsize]{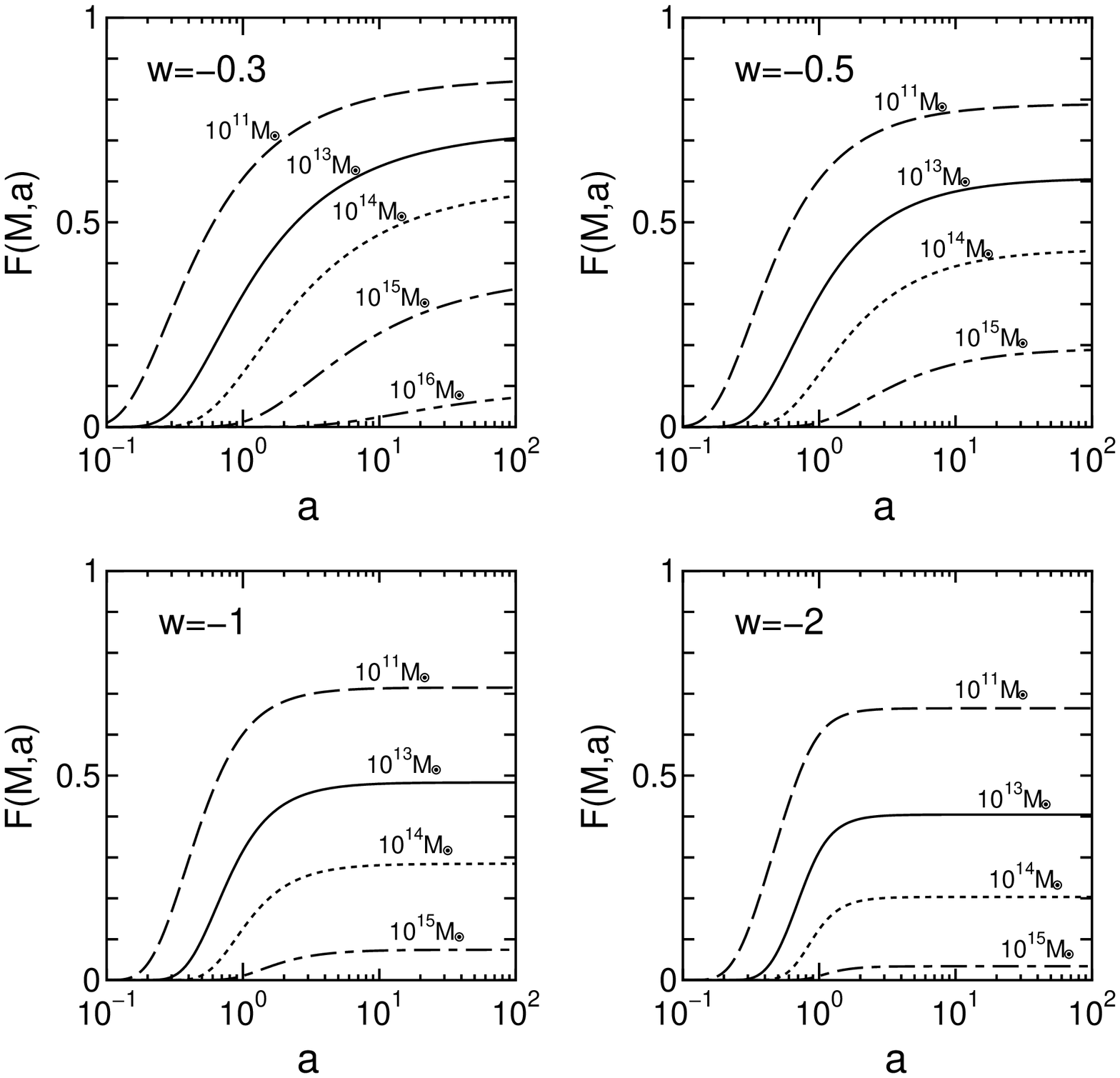}
\caption{
The fraction of collapsed objects with mass greater than $M$ as a function
 of a.}
\label{ps}
\end{figure}

The fraction of collapsed objects with mass greater than $M$ at the scale
 factor $a$ is given by \cite{ps74}
\beq
  F(M,a) = \frac{2}{\sqrt{2 \pi} \delta_M(a)} \int_{\delta_c(a)}
 ^{\infty} d \delta e^{- \delta^2 / 2 \delta_M^2 (a)},
\eeq 
where $\delta_M(a)$ is the mass fluctuation at $a$, $\delta_M(a)=
 \delta_M D(a)/D_0$.
The critical density $\delta_c(a)$ is the linear perturbation when
 the sphere collapses at $a$. 
In Fig.\ref{ps}, we show $F(M,a)$ as a function of the scale factor
 with $M=10^{11}
 M_\odot$ (dash), $M=10^{13} M_\odot$ (solid), $M=10^{14} M_\odot$ (dot),
 $M=10^{15} M_\odot$ (dot-dash) and $M=10^{16} M_\odot$ (dot-dot-dash).
For larger $w$ the objects with the larger mass $M \simg 10^{15}
 M_{\odot}$ can form, while for smaller $w$ the growth of density
 fluctuation is suppressed and the mass fraction becomes constant
 in the near future.

\section{Summary}

In this paper, we have classified the future of the universe with dark energy with 
various equation of state. Moreover we have investigated the future 
structure formation of the universe. 

We have found that the phase diagrams of the universe in 
$\Omega_{M} -\Omega_{X}$ plane 
are drastically different between $w \geq -1/3$ and $w < -1/3$.
Infinitesimally small change of $w$ from $w=-1/3$ to $w<-1/3$ moves 
regions of ``expands forever'' and ``recollapses eventually'' to
completely different places.  Moreover, for $w < -1/3$, there is a 
region of ``no big bang'', which does not exist for  $w \geq -1/3$. 
On the other hand, phase diagrams of  $-1/3 > w \geq -1$ or
$w < -1$ are about same, except the ``expands forever'' region  of the
former is replaced with the ``big rip'' region of the latter.  

This discontinuity at $w = -1/3$ can be seen in conformal diagrams
too such that the event horizon appears for $w < -1/3$, but $w \geq -1/3$. 
The size of the event horizon grows, remains constant and decreases
for $-1 < w < -1/3$, $w=-1$, and $w<-1$, respectively. 

Concerning the structure formation, we have found the linear evolution
asymptotically converges to a constant value  for $w<0$ and obtained 
the value in the limit of $a \rightarrow \infty$ as a function of $w$.  

However, it is not always the case that these linear perturbations
eventually turn into collapsed objects.  We have classified the fate of the
overdense regions into three categories, i.e., ``monotonously
expands'', ``collapses'', and ``expands-contracts-re expands''.
And we have shown each region in $w - \Delta_0$ plane, where
$\Delta_0$ is the ratio of density of the spherical region to
background density at the present epoch.   We have also found 
the largest structure in the future universe as a function of $w$.  

\ack
This work was supported in part by a Grant-in-Aid for Scientific 
Research (Nos.15740152 and 14340290) from the Japan Society for the Promotion of
Science.

%%%%%%%%% references %%%%%%%%%%%%%%%%%%%%%%%%%%%%%%

\end{document}